\newcommand{\beq}{\begin{equation}}
\newcommand{\eeq}{\end{equation}}
\newcommand{\bea}{\begin{eqnarray}}
\newcommand{\eea}{\end{eqnarray}}

\newcommand{\gsim}{\lower.7ex\hbox{$\;\stackrel{\textstyle>}{\sim}\;$}}
\newcommand{\lsim}{\lower.7ex\hbox{$\;\stackrel{\textstyle<}{\sim}\;$}}

\newcommand{\mrm}{\mathrm}

\documentclass[aps,prev,twocolumn,preprintnumbers,floatfix,nofootinbib]{revtex4-1}
\usepackage{graphicx}
\usepackage{epstopdf}
\usepackage{mathrsfs}
\usepackage{amssymb}
\usepackage{verbatim}
\usepackage{color}
\usepackage{multirow}
\usepackage{amsmath}
\usepackage{subfig}
 \usepackage{slashed} 
\usepackage{float}

\def\stacksymbols #1#2#3#4{\def\theguybelow{#2}
    \def\vp{\lower#3pt}
    \def\sp{\baselineskip0pt\lineskip#4pt}
    \mathrel{\mathpalette\intermediary#1}}

\def\intermediary#1#2{\vp\vbox{\sp
     \everycr={}\tabskip0pt
     \halign{$\mathsurround0pt#1\hfil##\hfil$\crcr#2\crcr
              \theguybelow\crcr}}}

\def\Ms{M_{SUSY}} 
\def\MG{m_{3/2}}
\def\MP{M_{Pl}}

\def\be{\begin{equation}}
\def\ee{\end{equation}}
\def\bea{\begin{eqnarray}}
\def\eea{\end{eqnarray}}

\def\sp{\;\;\;,\;\;\;}

\def\mrm{\mathrm}

\def\lsim{\raise0.3ex\hbox{$\;<$\kern-0.75em\raise-1.1ex\hbox{$\sim\;$}}}
\def\gsim{\raise0.3ex\hbox{$\;>$\kern-0.75em\raise-1.1ex\hbox{$\sim\;$}}}

\def\inbar{\,\vrule height1.5ex width.4pt depth0pt}

\def\IC{\relax\hbox{$\inbar\kern-.3em{\rm C}$}}
\def\IQ{\relax\hbox{$\inbar\kern-.3em{\rm Q}$}}
\def\IR{\relax{\rm I\kern-.18em R}}
 \font\cmss=cmss10 \font\cmsss=cmss10 at 7pt
\def\IZ{\relax\ifmmode\mathchoice
 {\hbox{\cmss Z\kern-.4em Z}}{\hbox{\cmss Z\kern-.4em Z}}
 {\lower.9pt\hbox{\cmsss Z\kern-.4em Z}}
 {\lower1.2pt\hbox{\cmsss Z\kern-.4em Z}}\else{\cmss Z\kern-.4em Z}\fi}

\def\comment#1{}
\def\to{\rightarrow}

\def\u1x{U(1)_X}
\newcommand{\nc}{\newcommand}
\nc{\LL}{L}
\nc{\vv}{\tilde{v}}
\nc{\ccdot}{\!\cdot\!}
\nc{\gsm}{G_{SM}}
\nc{\vfive}{\mathbf{5}\oplus\mathbf{\overline{5}}}
\nc{\vten}{\mathbf{10}\oplus\mathbf{\overline{10}}}
\nc{\zhol}{Z^{\rm hol}}
\nc{\xfb}{\,{\rm fb}}

\begin{document}

%\wideabs{
%\begin{flushright}
%
%
%\end{flushright}

\preprint{LPT--Orsay 16-93}
\preprint{CPHT-RR070.122016}

\vspace*{1mm}

\title{A Minimal Model of Gravitino Dark Matter}

\author{Karim Benakli$^{a}$}
\email{kbenakli@lpthe.jussieu.fr}
\author{Yifan Chen$^{a}$}
\email{yifan.chen@lpthe.jussieu.fr}
\author{Emilian Dudas$^{b}$}
\email{emilian.dudas@cpht.polytechnique.fr}
\author{Yann Mambrini$^{c}$}
\email{yann.mambrini@th.u-psud.fr}

\vspace{0.1cm}
\affiliation{
${}^a$ 
LPTHE, Sorbonne Universit\'es, UPMC Univ Paris 06, CNRS, UMR 7589, 4, Place Jussieu, F-75005, Paris, France}
\affiliation{
${}^b $ CPhT, Ecole Polytechnique, 91128 Palaiseau Cedex, France 
}
\affiliation{
${}^c $ Laboratoire de Physique Th\'eorique, CNRS -- UMR 8627, \\
Universit\'e de Paris-Saclay 11, F-91405 Orsay Cedex, France}

\begin{abstract} 

Motivated by the absence of signals of new physics in both searches for new particles at LHC and for a Weakly Interacting Massive Particle (WIMP) dark matter candidate, we consider a scenario where supersymmetry is broken at a scale above the 
reheating temperature.  The low energy particle content consists then only in Standard Model states and a gravitino. 
We investigate the possibility that the latter provides the main component of dark matter through the annihilation of thermalized Standard Model particles. 
  We focus on the case where 
its production through scattering in the thermal plasma is well approximated by the non-linear supersymmetric effective
 Lagrangian of the associated goldstino and identify the parameter space allowed by the cosmological constraints, 
 allowing the possibility of large reheating temperature compatible
 with leptogenesis scenarios, alleviating the so called "gravitino problem". 
 
\end{abstract}

\maketitle
%}

%%%%%%%%%%%%%%%%%%%%%%%%%%%%%%%%%%%%%%%%%%%%%%%%%%%%%%%%%%%%%%%%%
%%%%%%%%%%%%%%%%%%%%%%%%%%%%%%%%%%%%%%%%%%%%%%%%%%%%%%%%%%%%%%%%%
%%%%%%%%%%%%%%%%%%%%%%%%%%%%%%%%%%%%%%%%%%%%%%%%%%%%%%%%%%%%%%%%%

\maketitle

%% The arXiv's use of hypertex conflicts with revtex4's use of
%% \tableofcontents in single column format. To avoid this problem,
%% Include a file OOREADME.xxx with the word nohypertex in it when
%% you submit to the arXiv.
%\tableofcontents

%\section{Introduction}\label{sec:introduction}
\setcounter{equation}{0}

%%%%%%%%%

%%%%%%%%%%%%%%%%%%%%%%%%%%%%%%%%%%%%%%%%%%%%%%%%%%%%%%%%%%%%%%%%%%%%%%

\section{Introduction}

Among the possible hidden symmetries of Nature, supersymmetry is of the most appealing. It endows a candidate for a fundamental theory of Nature with a better ultraviolet behavior. If it is realized at low energy,  it allows to address the hierarchy problem of the electroweak symmetry breaking sector, provides a dark matter candidate when R-parity is conserved and allows unification of the gauge couplings. However, the benefits of this aesthetically attractive scenario need to be questioned in the light of increasing tensions with experimental data. On one hand, there is no sign for new physics in the searches at LHC implying strong constraints on the parameter spaces of the models. On the other hand, the negative results in direct and indirect  searches for WIMPs are closing the window of parameters corresponding  to a neutralino dark matter. This  motivates to push further up the scale of supersymmetry breaking extending the energy range of  validity of the Standard Model. We consider here this possibility within a peculiar cosmological scenario.

Weakly Interacting Massive Particle (WIMP) \cite{wimp} and Freeze-In Massive Particle (FIMP) \cite{fimp} are different theoretical frameworks that have been postulated for the production mechanisms of dark matter. Whereas WIMP dark matter are in equilibrium with other particles in the early universe, when the temperature drops below the dark matter mass, they freeze-out and then form the present relic abundance. In supersymmetric frameworks, the lightest neutralino with a mass around the weak scale is a natural WIMP candidate \cite{wimp}.
The recent analysis on dark matter detections, from the indirect methods by its annihilation products
(FERMI \cite{FERMI}, HESS \cite{HESS}, AMS \cite{AMS}) or its direct detection processes through the measurements of nuclear recoils 
(LUX \cite{LUX}, PANDAX \cite{PANDAX} or XENON100 \cite{XENON})
still did not find any significative signals. However, the sensitivity reached by the different groups begins to exclude
large parts of the parameter space predicted by simple WIMPly extensions of the Standard Model (Higgs-portal \cite{Hportal}, 
Z-portal \cite{Zportal}, even Z' portal \cite{Zpportal}). At the supersymmetric level, the auhtors of \cite{Badziak:2017the} showed
that the well tempered neutralino, which was
the most robust supersymmetric candidates during the last years in now  severily constrained by the last results of LUX \cite{LUX}. Wino--like neutralino is also severely constrained by the latest indirect detection searches released by the
FERMI collaboration \cite{FERMI}. A nice review of the status of dark mater SUSY searches in different supersymmetric
scenario with neutralino dark matter can be found in \cite{Baer:2016ucr}. In any case, the recent prospects exposed by the
LUX \cite{LUXfuture} and FERMI \cite{FERMIfuture} collaboration showed that the WIMP paradigm should be excluded 
(or discovered),  for dark matter masse below 10 TeV in the present generation of detectors.

The lack of experimental signals motivates investigations of other production mechanisms with weaker couplings. FIMP dark matter is an alternative using couplings between the dark matter and Standard Model particles suppressed by a much higher scale than the weak scale. Thus they are not in equilibrium with other particles in the early universe and never reach equilibrium among themselves, their yields keeping growing as temperature drops. One can distinguish three cases of FIMPs: (i) Decay products of some heavier particles in equilibrium; (ii) Final products of infrared (IR) [$M_{DM}$] dominated processes; (iii) Final products of ultraviolet (UV) [$T_{RH}$] dominated ones. Gravitinos produced by the decay of Next to Lightest Supersymmetric Particles (NLSP) \cite{Cheung:2011nn} belongs to the first class. Because the production is dominated at low temperatures, the final yields are largely independent of the reheating temperature. This scenario has a constraint from Big Bang nucleosynthesis (BBN) since
 the late decay of NLSP influences the nucleosynthesis mechanisms strongly constrained by the observed abundance of D and $\mathrm{^4He}$ in the Universe \cite{Pradler:2006qh}. The IR dominated productions usually correspond to renormalizable operators or 2 $\rightarrow 1$ processes \cite{fimp}. These are most efficient when the temperature is near the FIMP mass. Thus the yields are not dependent on the reheating temperature. On the other hand, non-renormalizable operators usually lead to UV dominated production and the final results are highly dependent on the reheating temperature.

Gravitino is an universal prediction of local supersymmetry models. 
Its role in cosmology depends on its abundance and its lifetime. Even if in some non-minimal scenario 
it can be non-thermally produced at the end of inflation during preheating due to fine-tuned coupling to the inflaton,
 the amount expected is model dependent and can be small \cite{gravitinoreheat,Ema:2016oxl}. We will however not consider this possibility
  in this work and will instead focus on the case of thermal production. 
The gravitinos are produced by scattering of Standard Model states in the thermal plasma after reheating or through the
 decay of the NLSP. Within the assumption that the reheating temperature is lower  than the mass of all the supersymmetric 
 particles, we are left with only the former possibility.  
 However, the standard scenario of gravitino dark matter suffers from different cosmological difficulties
which are referred to collectively as the "cosmological gravitino problems": i) the late decaying superpartners can 
strongly affect the Big Bang Nucleosynthesis \cite{Pradler:2006qh}, and ii) if thermalized, the relic gravitinos produced  overclose the Universe
for $m_{3/2} \gtrsim 1$ keV \cite{Pagels:1981ke}, making it difficult to be a warm dark matter candidate if one also takes in consideration large scale structure
formations, Tremaine-Gunn bound or Lyman $\alpha$ constraints \cite{Tremaine:1979we}.
 We will show that in the high scale Supersymmetry framework we propose, where the gravitino is directly produced from the thermal bath scattering, these two issues do not hold anymore. As a consequence, our analysis leads naturally to a prediction of possible large reheating temperature $T_{RH}$, usually favored by inflationary or leptogenesis scenarios.
 
 \noindent
 The paper is organized as follows: we settle the framework of our model in section II, insisting on the fundamental mass scales entering in the analysis, before building the effective Lagrangian and computing the cosmological observables in section III. We then conclude in section IV.

\section{Supersymmetry breaking and the reheating temperature}

\noindent
We review in this section the different scales relevant for our analysis, 
and their origins: the SUSY breaking scale, the soft mass terms, the messenger scale and the gravitino mass.

\begin{enumerate}

\item{\bf The supersymmetry breaking parameters: } 

We denote by $F$ the order parameter for supersymmetry breaking, a generic combination of auxiliary $F$ or $D$ terms vacuum expectation values . It corresponds to  a spontaneous  breaking, thereof it implies the existence of a Goldstone fermion, the goldstino $G$. The super-Higgs mechanism at work  leads to a mass for the gravitino which value at present time reads \cite{deserzumino}:
\beq
\MG= \frac {F}{\sqrt{3} \MP},
\eeq
in which $M_{Pl}$ is the reduced Planck mass. The breaking is mediated to the visible sector through messengers lying at a scale $\Lambda_{mess}$. This leads to soft-terms of order $\Ms$:
\beq
\Ms = \frac {F}{\Lambda_{mess}} .
\eeq

We shall assume for simplicity that all the masses of sparticles, squarks, sleptons, gauginos and higgsinos as well as all the new scalars in the extended Higgs sector are at least of the order of the scale of supersymmetry breaking $\Ms$. These particles are thus decoupled at reheating time $T_{RH}$. Below $\Ms$, the particle content is the Standard Model (SM) (with possibly right-handed neutrinos) and the goldstino. 
How realistic is this assumption in explicit supersymmetry breaking models is a model-dependent question. In O'Rafeartaigh models of supersymmetry breaking, the partner of the goldstino, the sgoldstino $\tilde G$, is usually massless at tree-level. Quantum corrections are however expected to fix its mass to be one-loop suppressed with respect to the supersymmetry breaking scale $m_{\tilde G}^2 \sim \frac{g^2}{16 \pi^2} F$, that has to be above the reheating temperature for our model to be self-consistent.  In string effective supergravities it is also often the case for the sgoldstino to be light,  with mass of the order of the gravitino mass \cite{scrucca}.  This is also however a model-dependent statement; this can be avoided in models with a large Riemann curvature in the Kahler space \cite{kalloshlinde}.    
On the other hand, asking for $m_{3/2} \ll M_{SUSY}$ implies
\beq
{\Lambda_{mess}}\ll \MP
\eeq
and in the energy range under consideration the Renormalisation Group Equations (RGEs) are those of the SM. 
In particular for a Higgs boson of 126 GeV, it leads to a vanishing of the quartic coupling at scales of order 
$2 \times 10^{10}$ GeV to $3 \times 10^{11}$ GeV depending on the assumption on the degeneracy of superparticles 
soft masses, the exact value of the top mass and the strong interaction gauge coupling 
(see for instance \cite{Bagnaschi:2014rsa}). We then considered
\beq
\Ms \lesssim \{10^{10} -10^{11} \} {\rm GeV} .
\label{Eq:msusy}
\eeq
Supersymmetry breaking scales above this value can be achieved by a modification of the RGEs through introduction of new light particles. We shall not discuss these cases in details here, the generalisation being straightforward.

\item{\bf The cosmological parameters:}

The cosmological history described here starts after the Universe is reheated. Some assumptions are made for this epoch: (i)  The 
reheating temperature $T_{RH}$ is small enough to not produce superpartners of the Standard Model particles, thereof $T_{RH}
 \lesssim \Ms$ (ii) in the reheating process goldstinos are scarcely produced. This second condition is a constraint of the nature of the 
inflaton, its scalar potential and the branching ratios in its decay. A discussion of the production of goldstinos at the end of inflation can
 be found for example in \cite{gravitinoreheat} .

We consider that the dark matter gravitino interactions are well approximated by the helicity $\pm 1/2$ components. This is true in 
virtue of the equivalence theorem if the energy $E$ of the gravitinos is much bigger than their mass.
The enhancement of the interactions of the (longitudinal component of the) light
gravitino is a direct consequence of 
the equivalence theorem between the goldstino and the longitudinal component of the
gravitino $\Psi_\mu \to \left(\frac{1}{m_{3/2}}\right) \partial_\mu G $, as
discussed for the first time in \cite{Fayetbis}.
Approximating the former by the
 temperature $T$ of the SM particles in equilibrium leads  to the mass hierarchies that defines the self-consistency of our setup:
\beq
~ ~ ~ ~ ~ ~ ~ ~ \MG \ll  T_{RH} \lesssim \Ms \lesssim \sqrt{F} \lesssim {\Lambda_{mess}}\ll \MP %\nonumber
\eeq

Note that our bound on the reheating temperature is compatible with thermal leptogenesis. In fact, a lower bound of  the reheating temperature is obtained when the latter is identified  with the mass of the lightest right handed neutrino. It is at most of order $10^9$ GeV but can be lower depending on the assumptions on the initial abundance and the mass hierarchies of the neutrinos  (see for example \cite{Davidson:2002qv}).

\end{enumerate}

\section{Goldstino Dark Matter}

\subsection{Effective   goldstino interactions}

Under the assumption $\MG \ll E \sim T$ discussed above, the gravitino interactions with SM fields are dominated by the helicity
 $\pm1/2$ components. Moreover, for $E \sim T \lesssim T_{RH} \lesssim \Ms$, these are described by a non-linear realization 
 of supersymmetry in all the observable SM sector, since we will consider all superpartners to be heavy and therefore not accessible 
 in the thermal bath after reheating\footnote{Reheating temperature below superpartner masses was proposed and investigated in particular in \cite{Giudice:2000ex} and \cite{monteux}. The novelty in our case is that we consider high-scale supersymmetry, so our reheating 
 temperature is much higher compared to these references.}.  The leading order goldstino-matter interactions  can be divided into two types of 
 contributions: universal \cite{wessbagger} and non-universal ones \cite{bfz,ks,tg}. We will restrict our analysis to the former,
 which corresponds to the minimal couplings expected from the low energy 
 theorem\footnote{As we will see, our result will not depend drastically on this hypothesis}. 
 Their construction starts by defining  a "vierbein" \cite{va}

\begin{equation}
e_m^a \ = \ \delta_m^a -\frac{i}{2 F^2} \partial_m G \sigma^a {\bar G} +  \frac{i}{2F^2} G \sigma^a \partial_m {\bar G} \ , 
\label{va1}
\end{equation}
that under a supersymmetry transformation of parameter $\epsilon$ transform as a diffeomorphism
in general relativity
\begin{equation}
\delta e_m^a \ = \ \partial_m \xi^n e_n^a + \xi^n \partial_n e_m^a \ , 
\label{va2} 
\end{equation}
where $\xi^n = \frac{i}{F} e_a^n ( \epsilon \sigma^a {\bar G} - G \sigma^a {\bar \epsilon})$. 
The couplings to matter in this original geometrical prescription follows therefore the standard coupling to matter of a metric tensor built out from the vierbein $g_{mn} = \eta_{ab }e_m^a e_n^b $.
The corresponding goldstino-matter effective operators are consequently of dimension eight and take the form: 
\begin{equation}
L_{2G} = \frac{i}{2F^2}(G\sigma^\mu\partial^\nu\bar{G} - \partial^\nu G\sigma^\mu\bar{G}) T_{\mu\nu},
\end{equation}
%\noindent
where $G$ is the goldstino field and $T_{\mu\nu}$ is the energy momentum tensor of the SM matter fields. The  energy momentum tensor is given by:

\bea
&&
T_{\mu\nu} = + \eta_{\mu\nu}\tilde{L} \nonumber
\\
&&+
[\sum_{f}(-\frac{i}{4}D_\mu\bar{\psi}_f\bar{\sigma}_\nu\psi_f + \frac{i}{4}\bar{\psi}_f\bar{\sigma}_\nu D_\mu\psi_f)
- D_\mu H D_\nu H^\dagger\nonumber\\ 
&&+ \sum_{SM group} \frac{1}{2} F^{a\xi}_{\mu} F^a_{\nu\xi} + (\mu \leftrightarrow \nu)] .
\eea

\noindent
The scalar potential and mass terms for scalar and fermions appear in the first term. After the contraction between $\eta_{\mu\nu}$ and $G\sigma^\mu\partial^\nu\bar{G}$, the on-shell production of two goldstinos give a cross section proportional to $\MG^2$. As $\MG$ is much smaller than $T_{RH}$, these contributions can be neglected, as we will see later. Then the $2\rightarrow2$ scatterings for the goldstino production is dominated by the following operators\footnote{See the appendix for the expression of these operators in 4-component Dirac spinors and $\gamma$-matrices notation.}
\bea
&&
\frac{i}{2F^2} (G\sigma^\mu\partial^\nu\bar{G} - \partial^\nu G \sigma^\mu  \bar{G}) ( \partial_\mu H \partial_\nu H^\dagger +
\partial_\nu H \partial_\mu H^\dagger) , \nonumber
%\label{Eq:hprod}
\\
&&
\frac{1}{8F^2} (G\sigma^\mu\partial^\nu\bar{G} - \partial^\nu G \sigma^\mu \bar{G} )  \times  \nonumber
% \label{Eq:fprod} 
\\
&&
 (\bar{\psi}\bar{\sigma}_\nu \partial_\mu\psi + \bar{\psi} \bar{\sigma}_\mu \partial_\nu\psi 
 -\partial_\mu {\psi} \bar{\sigma}_\nu \psi -   \partial_\nu {\psi} \bar{\sigma}_\mu \psi ) ,
 \nonumber
 \\
&&
\sum_{a} \frac{i}{2F^2} (G\sigma^\xi\partial_\mu\bar{G} -  \partial_\mu G \sigma^\xi  \bar{G} ) F^{\mu\nu a}F^a_{\nu\xi},
\label{Eq:gprod}
\eea

\noindent
where $h$, $\psi$ and $F^a_{\nu\xi}$ stand for a complex scalar (Higgs doublet), gauge bosons and two-component fermions (quarks and leptons), respectively. 
\noindent
Another way to describe the two goldstinos interactions to matter is to replace the superpartner soft mass terms by couplings between the goldstino superfield and the matter superfield multiplets. One can integrate out the heavy (superpartner) components and eliminate them as a function of the light degrees of freedom : the SM fields and goldstino. This leads to an effective low-energy theory where the incomplete multiplets are described in terms of constrained superfields \cite{ks,dgglp}. The kinetic terms of the sparticles will then lead to dimension-eight operators  containing two goldstinos and two SM fields that generically differ from the ones computed from the low-energy theorem couplings \cite{bfz}. For the gauge  and the SM fermion sectors, the resulting cross sections only differ in the angular distribution and numerical constants, whereas the energy dependence is the same as for the low-energy theorem couplings. 

Since the masses of the superpartners are of order $M_{SUSY} < \sqrt{F}$, one can worry about effective operators generated after decoupling heavy superpartners, with larger coefficients. In particular, there can be dimension-eight operators proportional to $1/M_{SUSY}^4$ and 
$1/M_{SUSY}^2 F$, that would be dominant over the universal couplings we use in our paper. This issue was investigated in the first reference
in \cite{dgglp}, where it was shown that starting from MSSM only dimension-eight R-parity violating couplings of this type are generated.
The reason for this is the following: Integrating out heavy superpartners (without R-parity violation) 
leads to
factors of $1/M_{SUSY}$ from the propagators 
(the square of them for scalar
superpartners). The leading interactions of goldstino to matter, on the other hand,
through the soft terms, are proportional to $M_{SUSY}/F$ (the square for the scalar
superpartners). As a result, the factors of $M_{SUSY}$ cancel out leaving generically 
dimension-eight operators suppressed by $1/F^2$.
The effect of the R-parity violating couplings on the gravitino production was investigated more recently in \cite{monteux}.

%%%%%%%%%%%%%%%%%% RELIC DENSITY   %%%%%%%%%%%%%%%%%%%%%%%%%%%%%%%%%

\subsection{Computation of the gravitino relic density}

\subsubsection{The framework}
\noindent
Contrarily to the weakly interacting neutralino, the gravitino falls in the category of feebly interacting dark matter. Its interactions at high energies are governed by the helicity-1/2 component whose couplings are naturally suppressed by the supersymmetry breaking scale. In gravity mediated supersymmetry breaking the gravitino
is often heavier than the supersymmetric spectrum that it generates. As a consequence, if the gravitino is not sufficiently
heavy (ie below 30 TeV) it is a long-lived particle which  usually decay around the BBN epoch. 
This gives rise to the famous "gravitino problem" \cite{Ellis:1984er,GR}. In that case, in order to minimize the observable effects,
the gravitino density has to be small enough at the cost of an upper bound on the reheating temperature of the Universe
(see eg \cite{Moroi:1993mb}). On the other hand, if gravitino is the LSP, it can be a very good dark matter candidate,
either as stable or metastable particle, with lifetime much longer than the age of the Universe. 

The gravitino was in fact the first supersymmetric dark matter candidate ever proposed\footnote{TO be exact, P. Fayet in the Moriond proceeedings 1981 and 1982 \cite{Fayet} already proposed such an hypothesis} by 
Pagels and Primack \cite{Pagels:1981ke} and Khlopov and Linde \cite{Khlopov:1984pf}.
However, they showed that if thermalized, its mass is restricted to a window $m_{3/2} \lesssim$ 1 keV which place it 
in the hot scenario, nowadays in strong tensions by large scale formation constraints \cite{Kunz:2016yqy} or Tremaine Gunn bound ($m_{3/2} \gtrsim 400$ eV) \cite{Tremaine:1979we}.
Then, the authors of \cite{Moroi:1993mb} famously showed that the overabundance problem
can be avoided if, instead of thermalizing, the gravitino is produced through
scattering of gaugino with a reheating temperature below a critical value, depending
on the gaugino spectrum. They obtained 

\beq
\Omega_{3/2} h^2 \sim 0.3 \left( \frac{1 ~\mrm{GeV}}{m_{3/2}} \right) \left( \frac{T_{\mrm{RH}}}{10^{10}~\mrm{GeV}} \right) \sum_i c_i \left( \frac{M_i}{100~\mrm{GeV}} \right)^2, 
\label{Eq:omegaclassic}
\eeq

\noindent
where $c_i$ are coefficient of order one, and $M_i$ are the three gaugino masses. We clearly see from Eq.(\ref{Eq:omegaclassic})
that the density is settled by the reheating temperature. Lower limits on $M_3$ obtained by the non-observation of gluino at LHC set
(for a given gravitino mass) an $upper$ limit on reheating temperature to avoid overclosure of the Universe. These constraints are usually in tension with baryogenesis mechanisms \cite{Davidson:2002qv}, even if some interesting scenario with low reheating temperature ($T_{RH} \lesssim$ 100 TeV) can be found in \cite{Arcadi:2013jza}.

\noindent
Moreover, later on in \cite{Cheung:2011nn} it was shown that another contribution,
called "gravitino freeze in" is playing an important role. It corresponds to the decay of the superpartners while they are still in thermal equilibrium. Indeed, 
for a sufficiently large supersymmetric spectrum (as it seems to be observed at the LHC) the short lifetime of squarks or sleptons induced this process. The only way
to circumvent the overabundance, is to lower the reheating temperature $below$ the
supersymmetric spectrum to deal with the queue of the distribution. However, 
the origin of the gravitino is still the supersymmetric partners, through their decay.
A nice summary can be found in \cite{Hall:2013uga}. Adding the BBN constraints
give an upper bound on the gravitino mass of about 10 GeV \cite{Moroi:1995fs, Bolz:2000fu, moultaka}.

All the scenario discussed above made the hypothesis of thermal production of gravitino, through
supersymmetric partners in thermal equilibrium with the primordial plasma. And, this thermalisation hypothesis is the deep source of tension between the cosmological observables (density of dark matter, structure formation, BBN or leptogenesis) and the data. However, if for some reasons the supersymmetric 
breaking scale is above the reheating temperature {\it while still keeping it at high scale}, the SM superpartners will be too heavy to reach the thermal equilibrium. That solves naturally the preceding tension, but the issue of the gravitino production remains. Note also that for $T_{RH} \lesssim \Ms$ thermal corrections to the effective potential may become small enough to lead to only negilgeable displacements of the scalars vacuum expectation values from their late time values. This kind of high scale SUSY scenario can be originated easily and naturally in string inspired constructions (see \cite{bsb} and references therein for instance). 
A way to populate the Universe with gravitinos is through a $direct$ freeze-in from the thermal bath itself. In this new scenario, the gravitinos are produced at a rate smaller than the one corresponding to the expansion of the Universe, therefore they do not have time to reach the thermal equilibrium. It "freezes in" in the process to reach it as the strong suppression of the scattering cross sections 
 by the scale $F^2$  in Eq.(\ref{Eq:gprod}) prevents the gravitinos to be in thermal equilibrium with the Standard Model bath. That is this scenario we propose to confront with cosmological data.
 
\subsubsection{Gravitino production through freeze in}

 \noindent
 From the interaction generated through the Lagrangian Eq.(\ref{Eq:gprod}), one can compute
 the production rate $R = n_{eq}^2 \langle \sigma v \rangle$ 
 of the gravitino $\tilde G$, generated  by the annihilation of the standard model bath of density
 $n_{eq}$.
 The detail of the computation is developed in the appendix Eq.(\ref{Eq:rfinal}), and we obtain

 \beq
R = \sum_i n_{eq}^2 \langle \sigma v \rangle_i \simeq 21.65 \times \frac{T^{12}}{F^4}
\label{Eq:Rfinal}
\eeq

\noindent 
The Boltzmann equation for the gravitino density $n_{3/2}$  
can be written
 
 \beq
 \frac{dY_{3/2}}{dx} =\left( \frac{45}{g_* \pi} \right)^{3/2} \frac{1}{4 \pi^2} \frac{M_P}{m_{3/2}^5}x^4 R,
 \eeq

 \noindent
 with $x=m_{3/2}/T$, $Y_{3/2}=n_{3/2}/\bf{s}$, $\bf{s}$ the density of entropy and  $g_*$ is the effective number of degrees of freedom thermalized at the time of gravitino decoupling  (106.75 for the Standard Model). Here, we use the Planck mass $M_P= 1.2 \times 10^{19}$ GeV. We then obtain after integration
 
  \beq
 Y_{3/2}= \frac{21.65 M_\mrm{P}T_{RH}^7}{28 \pi^2 F^4} \left( \frac{45}{g_* \pi} \right)^{3/2}
 \simeq  3.85\times 10^{-3} ~\frac{M_\mrm{P} T_{\mrm{RH}}^7}{F^4}
 \label{Eq:omega0}
 \eeq
 
  \noindent
  The relic abundance 
  
  \beq
  \Omega h^2 = \frac{\rho_{3/2}}{\rho^0_c} = \frac{Y_{3/2}~s_0~ m_{3/2}}{\rho^0_c}
  \simeq 5.84 \times 10^8~ Y_{3/2} \left( \frac{m_{3/2}}{1~\mrm{GeV}} \right) 
  \eeq
    is then 
 \beq
 \Omega_{3/2}h^2 \simeq 
 0.11 \left( \frac{100 ~\mrm{GeV}}{m_{3/2}} \right)^3 \left( \frac{T_{\mrm{RH}}}{5.4 \times10^7 ~\mrm{GeV}} \right)^7
 \label{Eq:omega}
  \eeq
 
 \noindent

\noindent
 As we notice, the dependence on the reheating temperature is completely different from the case where the gravitino
 is produced through the scattering of the gaugino in Eq.(\ref{Eq:omegaclassic}). A similar behavior can be observed
 in SO(10) framework  \cite{Mambrini:2013iaa} or in extended neutrino sectors \cite{Giudice:2000ex} .
  All these models have in common that
 the production process appears at the beginning of the thermal history, and is then very mildly dependent on the hypothesis
 or the physics appearing {\it after} reheating. The reheating temperature is then a prediction of the model (for a given gravitino
 mass) once one applies the experimental constraints of WMAP \cite{WMAP} and PLANCK \cite{PLANCK}.
 Another interesting point, is that a look at Eqs.(\ref{Eq:omega0}) and
  (\ref{Eq:omega}) shows that even the dependance on the particle content is very mild. Indeed, due to the 
  large power $T^7_{\mrm{RH}}$,
 the total number of degrees of freedom, or even channels does not influence that much the final reheating temperature, 
 which is predicted to be around $10^8$ GeV for a gravitino with electroweak scale.
 Even the hypothesis of universal couplings \cite{wessbagger} or non-universal ones \cite{bfz,ks} will not affect 
 drastically our Eq.(\ref{Eq:omega}).

 \vspace{1cm}

\vspace{1cm}

 \begin{figure}
\centering
\includegraphics[width=0.80\columnwidth]{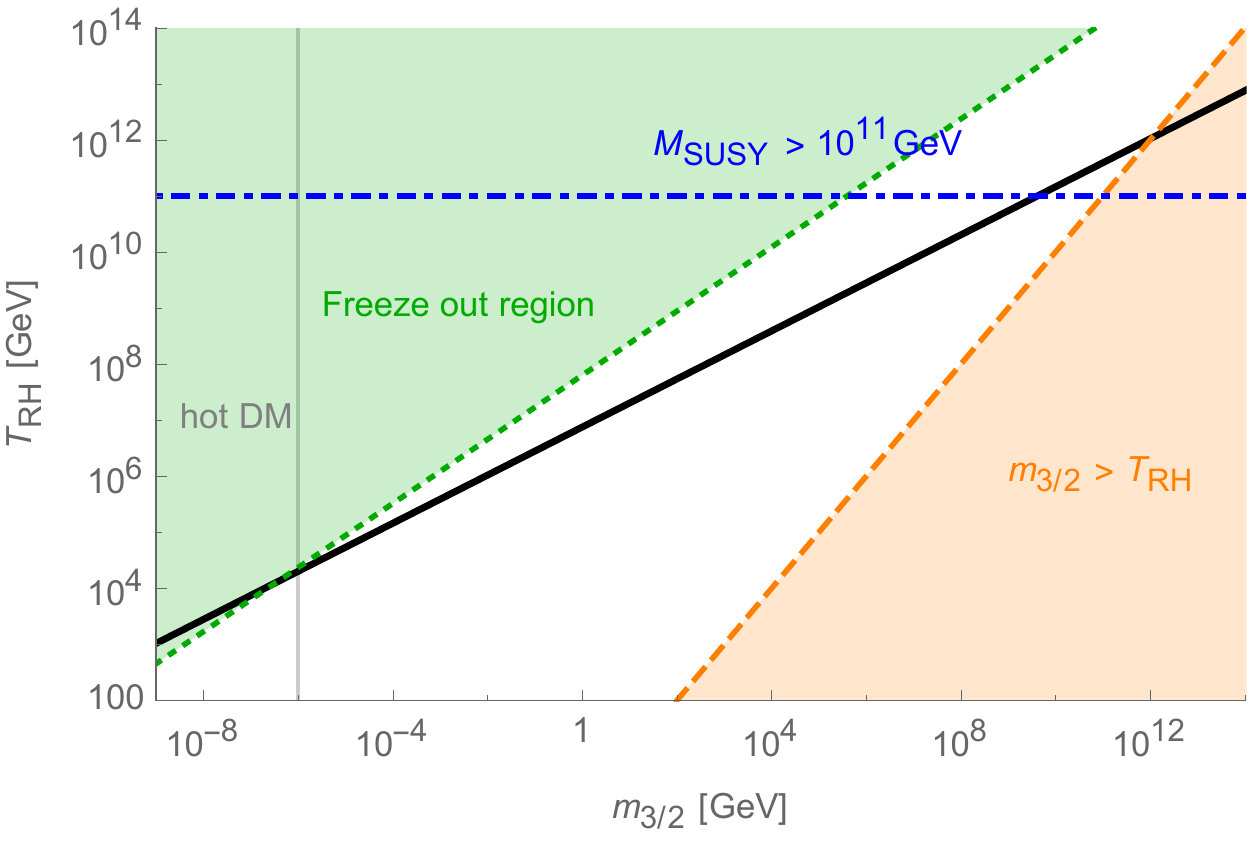}
\caption{
\footnotesize
{
Region in the parameter space ($m_{3/2};T_{RH}$) respecting the relic abundance constraint \cite{WMAP,PLANCK}
from Eq.(\ref{Eq:omega}). The points above the black line are excluded because gravitino would overclose the Universe.
The blue line constraint is from the Higgs mass with an observed value 125 GeV, which sets a upper limit for the scale of supersymmetry breaking (Eq.\ref{Eq:msusy}).
}
}
\label{Fig:mt}
\end{figure}

\noindent
Our result is plotted in Fig.(\ref{Fig:mt}) where we represent the parameter space allowed by the relic abundance
constraints $\Omega_{3/2} h^2 \simeq 0.12$ \cite{WMAP, PLANCK}. As we notice, there exist a large part of the parameter space 
allowed by cosmology, giving reasonable values of $T_{RH}\simeq 10^{5}-10^{10}$ GeV for a large range of gravitino masses 
MeV-PeV. 
The region below the orange (dashed) line is excluded as the gravitino would be too heavy to be produced by freeze--in
mechanism, whereas the region above the green (dotted) line corresponds to a freeze out scenario. In the latter region,
 the production cross section $\langle \sigma v \rangle$ is sufficiently high to reach the thermal equilibrium. 
 This occurs when  $n \langle \sigma v \rangle \gtrsim H(T_{RH}) \simeq T^2_{RH}/M_{Pl}$.
 A quick look at Eq.(\ref{Eq:Rfinal}) shows that such large cross section is obtained for high reheating temperature or small values of
 $F$ (and thus light gravitino), explaining the shape of the green region in Fig.(\ref{Fig:mt}). However, once the gravitino
 is in thermal equilibrium, its density is given by the classical Freeze Out  ($FO$) mechanism

 \beq
 \Omega^{FO}_{3/2} =\frac{n_{3/2} m_{3/2}}{\rho_c^0} ~ \Rightarrow ~ \simeq  0.1 \left( \frac{m_{3/2}}{180~\mrm{eV}} \right) 
 \eeq

\noindent
which corresponds obviously to the intersecting point in Fig.(\ref{Fig:mt}).

There exists potentially another non-thermal source of gravitino production: the decay of the NLSP.
Indeed, this contribution also exist in standard supersymmetric framework, through the relic abundance produced by the decay
of the NLSP (usually a sfermion $\tilde  f$) 
into $\tilde f \rightarrow G~ f $. This process being proportional
to $n_{\tilde f}^{eq}$, is highly
Boltzmann suppressed in our scenario where $T_{RH} \ll M_{NLSP}$. But there still exists some parameter space when NLSP are in equilibruim. Then the production of goldstinos is a combination of the decay of NLSP, QCD process and SM freeze in. An analysis in scenario with very low $T_{RH}$ 
($\lesssim$ GeV ) can be found in \cite{monteux}.

\begin{figure}
\centering
\includegraphics[width=0.80\columnwidth]{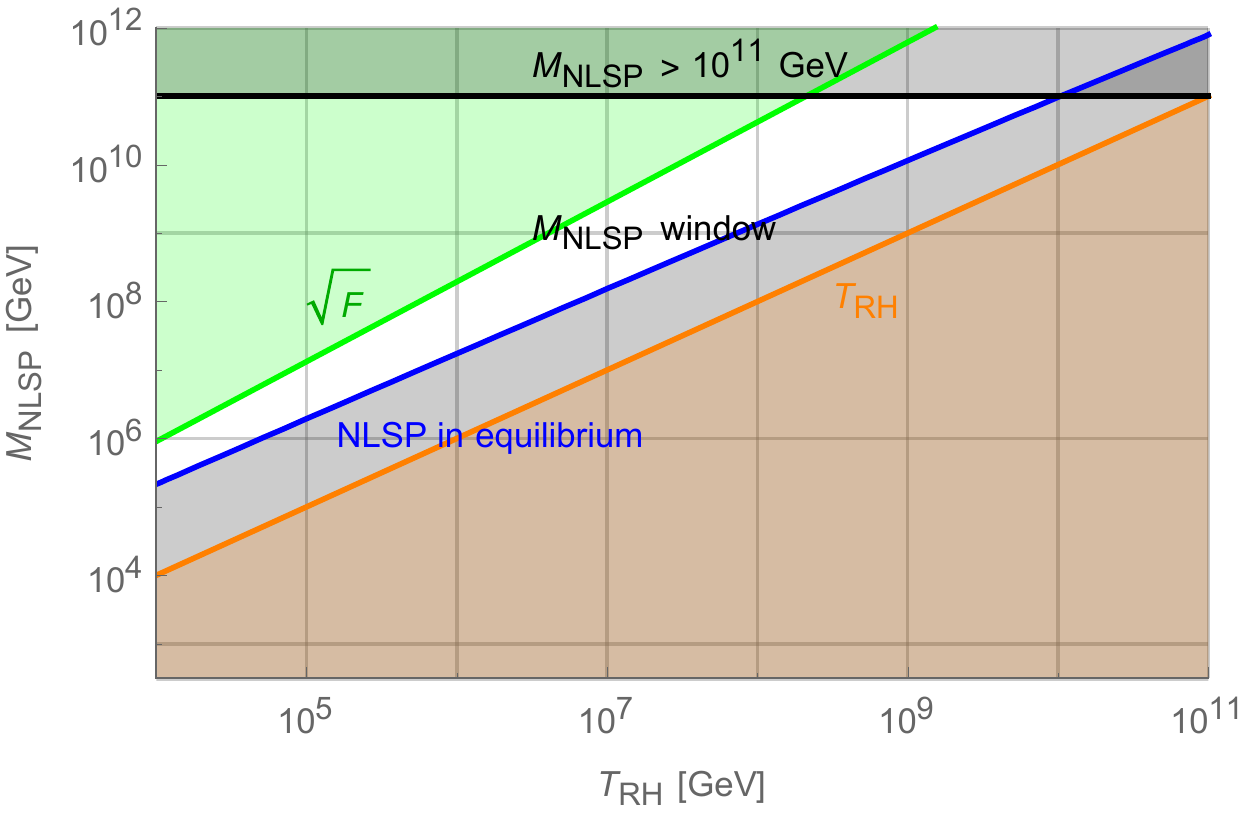}
\caption{The parameter space for $M_{NLSP}$. It must be lower than $\sqrt{F}$. The blue line corresponds to $n_{g}\langle\sigma v\rangle_{gg\rightarrow\tilde{g}\tilde{g}} = H$. Below the line, NLSP is still in equilibruim and can decay to gravitino.}
\end{figure}

%%%%%%%%%%%%%%%%%%%%%%%%%%%%%%%%%%%%%%%%%%%%%%%%%%%%%%%%%%%%%%%%%%%%%%%%%%%%

%\section{Observation potential}
\subsection{Comments on the R-parity violation operators}

R-parity violation operators can also be introduced in the  high-scale supersymmetry scenario discussed in this work. The corresponding operators involving goldstino fields include dimension-five operators as
\beq 
\frac{\mu_i}{F} l_I^i \sigma^\mu \bar{G} D^\mu h^j +  h.c,\\
\eeq 
dimension-six ones as \cite{rparity}
\beq 
\frac{iC^I}{F} \ \epsilon_{ij} (l_I^i\partial_\mu G)D^\mu h^j + h.c, \ ,  \label{r1}
\eeq 
and dimension-eight operators of the form
\begin{equation}
\frac{\lambda_{ijk}^{''}}{m_i^2 F } u_i d_j \Box (d_k G)  \ , \ \frac{\lambda_{ijk}^{'}}{m_i^2 F } q_i l_j \Box (d_k G) \ , \
\frac{\lambda_{ijk}}{m_i^2 F } l_i l_j \Box (e_k G)  \ ,  \label{r2}
\end{equation}
plus permutations. Here $\mu_i$ and $C$  are dimensionful and dimensionless coefficients respectively, $l_I$ are the three lepton doublets in the SM and $m_i^2$ are soft terms of the heavy superpartners that were integrated out. The 2 $\rightarrow$ 1 gravitino production through these operators will be suppressed at temperatures higher than the gravitino mass and only become important at late times, therefore do not need to be considered for the production of gravitino dark matter.

When R-parity is violated, the gravitinos are no more stable but can decay, giving rise to observable signatures. The latter are independent of the production mechanisms and the previous analysis in the literature apply to our case. The relevant operators can be derived from the above but should be written using the gravitino field. Since the heavy supersymmetric particles decouple in our case, the coefficients of the R-parity violating operators are not necessarily constrained from preserving baryon asymmetry as in previous studies \cite{RPV}.

\noindent 
However, one characteristic of our construction is that it allows for very heavy gravitino (above the PeV scale). Smoking gun signals like  tilde $G \rightarrow h \nu$  and $G \rightarrow \gamma \nu$ can be observable in telescope like Icecube for neutrino \cite{Aartsen:2013bka}, or the future Cerenkov Telescope Array (CTA) for the photon \cite{Consortium:2010bc}. In both case, a monochromatic high energy signal should be the signature of the gravitino decay, the spatial morphology distinguishing decaying dark matter (proportional to its density $\rho$) to annihilating one (proportional to $\rho^2$).

%%%%%%%%%
\section{Conclusion}

We considered the framework of high scale supersymmetry, where the scale
of superpartners $M_{SUSY}$ lies above the reheating temperature whereas the gravitino mass $m_{3/2}$ stays below. In this case, there still exist processes
which produce thermally gravitinos through scattering of the Standard Model particles
at the earliest time of reheating. Our result is well summarized by Fig.(\ref{Fig:mt}) and Eq.(\ref{Eq:omega}) where one can observe and understand the strong dependence of the relic abundance in the reheating temperature $T_{RH}$. Our result predicts large reheating temperature ($\sim 10^8$ GeV for a $\sim 100$ GeV gravitino). This scale pattern $m_{3/2} \ll T_{RH} \ll M_{SUSY}$ is common in some string models with high-scale supersymmetry breaking \cite{bsb} and opens new possibilities in model building.

\vskip.3in
\noindent
{\bf Acknowledgments}

\noindent We are grateful to I. Antoniadis, W. Buchmuller, T. Gherghetta, Y. Fazran, E. Kuflik, S. Pokorski and K. Turzynski for discussions. K.B and Y.M. acknowledge the support of the the European Research Council (ERC) under the Advanced  Grant Higgs@LHC (ERC-2012-ADG20120216-321133). The work of K.B and Y.C. is supported by the Labex ``Institut Lagrange de Paris'' (ANR-11-IDEX-0004-02,  ANR-10-LABX-63). The work of K.B is also supported by the Agence Nationale de Recherche under grant ANR-15-CE31-0002 ``HiggsAutomator''. E.D. acknowledges partial support from the Agence Nationale de Recherche grant Black-dS-String.  Y.M. wants to thank the Institute of Research for Fundamental Sciences hospitality in Teheran where part of this work has been completed and acknowledges the support by the Spanish MICINN's Consolider-Ingenio 2010 Programme under grant Multi-Dark {CSD2009-00064}, the contract  { FPA2010-17747}, the France-US PICS no. 06482, the LIA-TCAP of CNRS, the Research Executive Agency (REA) of the European Union under the Grant Agreement {PITN-GA2012-316704} (``HiggsTools'') and funding from the European Union's Horizon 2020 research and innovation programme under the Marie Sklodowska-Curie grant agreement No 674896. E.D and Y.M. would also like to acknowledge the support of the CNRS LIA (Laboratoire International Associ\'e) THEP (Theoretical High Energy Physics) and the INFRE-HEPNET (IndoFrench Network on High Energy Physics) of CEFIPRA/IFCPAR (Indo-French Centre for the Promotion of Advanced Research).

\noindent

%%%%%%%%%%%%%%%%%%%%%%%%%%%%%%%%%%%%%%%%%%%%%%%%%%%%%%%%%%%%%%%%%%%%%%%%%%%%%%%%%%%
\section*{Appendix}
%\label{Sec:appen}
%%%%%%%%%%%%%%%%%%%%%%%%%%%%%%%%%%%%%%%%%%%%%%%%%%%%%%%%%%%%%

\subsection*{Computing the gravitino production rate $R$}

We provide in this appendix the detail of the computation of the annihilation rate $n_{eq}^2 \langle \sigma v \rangle$.
Indeed, after symmetrization and by switching to four-component fermionic notation, one can extract from Eqs.(\ref{Eq:gprod}) the effective Lagrangian

\bea
{\cal L} \supset
&&
-\frac{i}{2 F^2} \left( \partial_\mu \bar{G} \gamma_\nu \frac{1+\gamma_5}{2} G - \bar{G} \gamma_\nu \frac{1+\gamma_5}{2}  \partial_\mu G \right)
\nonumber
\\
&&
\times \left(\partial^\mu H^\dag \partial^\nu H + \partial^\nu H^\dag \partial^\mu H \right)
\nonumber
\\
&&
+ \frac{1}{8 F^2}  \left( \partial_\mu \bar{G} \gamma_\nu \frac{1+\gamma_5}{2} G - \bar{G} \gamma_\nu \frac{1+\gamma_5}{2}  \partial_\mu G \right)
\nonumber
\\
&&\left(  \bar \Psi \gamma^\nu\frac{1+\gamma_5}{2} \partial^\mu \Psi -  \partial^\mu \bar \Psi \gamma^\nu\frac{1+\gamma_5}{2}  \Psi \right. \nonumber
\\
&&
\left. + \bar \Psi \gamma^\mu\frac{1+\gamma_5}{2} \partial^\nu \Psi -  \partial^\nu \bar \Psi \gamma^\mu\frac{1+\gamma_5}{2}  \Psi \right)
\nonumber
\\
&&
-\frac{i}{2 F^2} \left( \partial_\mu \bar{G} \gamma_\nu \frac{1+\gamma_5}{2} G - \bar{G} \gamma_\nu \frac{1+\gamma_5}{2}  \partial_\mu G \right) ~ F^{\mu\lambda  a}  F_{\lambda} ^{~ \nu a }
\nonumber
\\
\label{Eq:lagrangian4}
\eea

\noindent
with $G$ being the goldstino in a four-component Dirac fermion notation and  $H$, $\Psi$ and $F_{\mu \nu}$ are the Higgs field, Standard Model fermions and gauge field strength respectively. 
It becomes then straightforward to compute the averaged production rate $R$ for the process $1+2 \rightarrow 3 + 4 $ in the case
of early decoupling, when all the particles $i$ in the thermal bath, of temperature $T$, 
are relativistic ($m_i \ll T \Rightarrow E_i=p_i$):

\bea
R_i = n_{eq}^2 \langle \sigma v \rangle_i = \int f_1 f_2 d \cos \beta \frac{E_1 E_2 dE_1 dE_2}{1024 \pi^6} \int |{\cal M}|_i^2 d \Omega
\nonumber
\eea

\noindent
with $f_i = \frac{1}{e^{E_i/T}\pm1}$ for a fermionic (bosonic) distribution, $\beta$ is the angle between the colliding
particles 1 and 2 of energies $E_1$ and $E_2$ respectively in the laboratory frame, and $\Omega$ is the 
solid angle between the incoming particle 1 and outgoing particle 3 in the center of mass frame\footnote{See
refs \cite{Edsjo:1997bg} and \cite{fimp} for details.}. From Eq.(\ref{Eq:lagrangian4}) one can easily deduce

\bea
&&
|\bar {\cal M}|^2_h = \frac{s^4}{16 F^4} (\cos^2 \theta - \cos^4 \theta)
\\
&&
|\bar {\cal M}|_f^2= \frac{s^4}{256 F^4} (1 + cos \theta)^2 (1- 2 \cos \theta)^2
\\
&&
|\bar {\cal M}|^2_V= \frac{s^4}{128 F^4} (2 - \cos^2 \theta - \cos^4 \theta)
\eea

\noindent
for the scalar, fermionic and vectorial contribution respectively\footnote{The integration on the phase space 
should be treated with care,
noticing that the lorentz invariant $s=(P_1 + P_2)^2=2P_1.P_2=2 E_1 E_2(1-\cos \beta)$ in the laboratory frame
and $\int \frac{x^n}{e^x -1} = n ! \zeta(n+1)$.}.
The {\it total} averaged production rate, $n_{eq}^2 \langle \sigma v \rangle$ is then given by

\beq
R = \sum_i n_{eq}^2 \langle \sigma v \rangle_i = 4 n_{eq}^2 \langle \sigma v \rangle_h + 45 n_{eq}^2 \langle \sigma v \rangle_f + 12 n_{eq}^2 \langle \sigma v \rangle_V
\eeq

\bea
&&
n_{eq}^2 \langle \sigma v \rangle_h= \frac{48 \zeta(6)^2}{\pi^5 F^4}T^{12}=  \frac{48 \pi^7}{(945)^2F^4} T^{12}
\nonumber
\\
&&
n_{eq}^2 \langle \sigma v \rangle_f= \frac{72 \zeta(6)^2}{\pi^5 F^4} (\frac{31}{32})^2 T^{12}=  \frac{72 \pi^7}{(945)^2F^4} (\frac{31}{32})^2 T^{12}
\nonumber
\\
&&
n_{eq}^2 \langle \sigma v \rangle_V= \frac{264 \zeta(6)^2}{\pi^5 F^4}T^{12}=  \frac{264 \pi^7}{(945)^2F^4} T^{12}
\eea

\noindent
implying

\beq
R = \frac{6400~ [\zeta(6)]^2}{\pi^5 F^4}T^{12}=  \frac{6400~ \pi^7}{(945)^2F^4} T^{12} \simeq 21.65 \times \frac{T^{12}}{F^4}
\label{Eq:rfinal}
\eeq

%%%%%%%%%%%%%%%%%%%%%%%%%%%%%%%%%%%%%%%%%%%%%%%%%%%%%%%%%%%%%%%%%%%%%%%%%%%%%%%%%%%%%%%%%%%%%%

\end{document}